# Response to Comment on "Pairing and Phase Separation in a Polarized Fermi Gas"

Guthrie B. Partridge, Wenhui Li, Ramsey I. Kamar, Yean-an Liao, and Randall G. Hulet

Sept 27, 2006

**Zwierlein and Ketterle rely on subjective arguments and fail to recognize important differences in physical parameters between our experiment and theirs. We stand by the conclusions of our original report.**

Having failed to establish that trap anharmonicities or other objective mechanisms affect the conclusions of our study [1], Zwierlein and Ketterle [2] now make the more subjective assertion that our claims are not supported by the data presented. We stand by the statements and claims made in [1]. Furthermore, in emphasizing discrepancies between our results and theirs, Zwierlein and Ketterle ignore differences in physical parameters of the two experiments. Differences in aspect ratio [3] or temperature [4, 5], for example, can have profound effects on the results.